\documentclass[namedreferences,hyperref,optionalrh,solaromanenum]{spr-sola}

\usepackage{epstopdf}
\usepackage{graphicx}                    
\usepackage{color}                       
\usepackage{hyperref}

\usepackage{sidecap}
\usepackage{lscape}
\usepackage{tabularx}
\usepackage{rotating}
\usepackage{amsmath}

\renewcommand{\vec}[1]{{\mathbfit #1}}

\newcommand{\ppdr}[2]{\frac{\partial^2 #1}{\partial #2^2}}
\newcommand{\grad}{{\bf \nabla}}

\newcommand{\dl}{~{\mathrm d} l}

\newcommand{\avec}{\vec A}
\newcommand{\rvec}{\vec R}
\newcommand{\mvec}{\vec M}

\newcommand{\rr}{\vec r}
\newcommand{\xx}{\vec x}
\newcommand{\yy}{\vec y}

\newcommand{\rsun}{$R_\odot$}
\newcommand\norm[1]{\left\lVert#1\right\rVert}

\newcommand{\ip}{i^\prime}
\newcommand{\jp}{j^\prime}
\newcommand{\kp}{k^\prime}

\chardef\us=`\_

\begin{document}

\begin{frontmatter}

\title{Improved Tomographic Reconstruction of 3D Global Coronal Density from STEREO/COR1 Observations }

%
\author[addressref={aff1,aff2},corref,email={tongjiang.wang@nasa.gov}]{\inits{T.J.}\fnm{Tongjiang}~\lnm{Wang}\orcid{0000-0003-0053-1146}}
\author[addressref={aff2}]{\inits{C.N.}\fnm{C. Nick}~\lnm{Arge}\orcid{0000-0001-9326-3448}}
\author[addressref={aff1,aff2}]{\inits{S.I.}\fnm{Shaela I.}~\lnm{Jones}\orcid{0000-0001-9498-460X}}

\address[id=aff1]{Department of Physics, Catholic University of America, 620 Michigan Avenue NE, Washington, DC 20064, USA}
\address[id=aff2]{NASA Goddard Space Flight Center, Code 671, Greenbelt, MD 20770, USA}

%
\runningauthor{Wang et al.}
\runningtitle{Improved 3D Coronal Density Reconstruction}

 \begin{abstract}
 Tomography is a powerful technique for recovering the three-dimensional (3D) density structure of the global solar corona. In this work, we present an improved tomography method by introducing radial weighting in the regularization term. Radial weighting provides balanced smoothing of density values across different heights, helping to recover finer structures at lower heights while also stabilizing the solution and preventing oscillatory artifacts at higher altitudes. We apply this technique to reconstruct the 3D electron density of Carrington Rotation (CR) 2098 using two weeks of polarized brightness (pB) observations from the {\it inner coronagraph} (COR1) onboard spacecraft-B of the twin {\it Solar Terrestrial Relations Observatory} (STEREO), where the radial weighting function is taken as the inverse intensity background, calculated by averaging all the pB images used. Comparisons between density distributions at various heights from the tomography and magnetohydrodynamics (MHD) simulations show good agreement. We find that radial weighting not only effectively corrects the oversmoothing effect near the inner boundary in reconstructions using second-order smoothing but also significantly improves reconstruction quality when using zeroth-order smoothing. Additionally, comparing reconstructions for CR 2091 from single-satellite data with that from multi-viewpoint data suggests that coronal evolution and dynamics may have a significant impact on the reconstructed density structures. This improved tomography method has been used to create a database of 3D densities for CRs 2052 to 2154, based on STEREO/COR1-B data, covering the period from 08 January 2007 to 17 September 2014.

 \end{abstract}

%

\end{frontmatter}

%

\section{Introduction}
The solar corona plays a vital role in solar wind formation, space weather forecasting, and understanding the Sun's magnetic field. In particular, knowledge of the three-dimensional (3D) electron density distribution is essential for enhancing observational diagnostics of coronal physical properties, interpreting impulsive events (e.g., coronal mass ejections, CMEs), and validating global magnetohydrdynamics (MHD) models of solar winds. For instance, tomographic 3D electron densities have been used to validate coronal magnetic field models, such as potential field source surface (PFSS) models \citep{kra14,kra16a,llov17}, Wang-Sheeley-Arge (WSA) solar wind models \citep{jon22}, and MHD coronal models \citep{vas08,kra14, llov22}. Based on tomographic 3D electron densities, \citet{fra03} used the {\it Ultraviolet Coronagraph Spectrometer} (UVCS) observations to determine the velocity distributions and outflow speeds of O$^{5+}$ ions in equatorial streamers. Additionally, tomographic 3D coronal density and temperature data have been used to reconstruct 3D coronal magnetic fields from Fe\,{\sc xiii} 10747 \AA\ polarization measurements, as well as to diagnose the thermal properties of coronal loops and their heating mechanisms \citep{mac20,mac22}.

The 3D electron density model is also critical for interpreting solar radio emissions, such as type II and type IV radio bursts generated by coronal eruptions. For example, knowledge of coronal electron density distributions enables direct estimates of the coronal magnetic field from split-band type II radio bursts \citep{ash17,ash19}. By analyzing a moving type IV radio burst observed at 80 MHz, resulting from second harmonic plasma emission in a CME leading edge, \citet{har16} determined the strength of the coronal magnetic field based on measured electron densities from coronagraph observations.  Moreover, accurately determining electron density distribution is essential for applying coronal seismology to estimate the Alfv\'{e}n speed and magnetic field strength in the corona using observed coronal MHD waves, such as pervasive propagating Alfv\'{e}nic waves \citep[e.g.][]{tom07,yang20}, CME-generated streamer kink waves \citep{chen10,chen11}, or global fast magnetoacoustic waves \citep[e.g.][]{kwon13}.

Inversions of electron density from white-light observations of the K corona have long been a classical problem in coronal physics. Since the K-coronal emission is optically thin and contributions come from electrons along the entire line of sight (LOS), extracting the electron density from a single 2D image of total brightness ($B$) or polarized radiance ($pB$) often requires assumptions about the distribution of electrons along the LOS. Historically, spherically symmetric models have been used for such inversions \citep[e.g.][]{vdh50,que02,wan14}. For instance, \citet{dec19} successfully inverted the 3D density of a coronal streamer by forward modeling observational data from two vantage points near quadrature, based on assumptions of the 3D structure of the streamer. 

However, tomography inversion offers a more robust method, as it does not rely on any assumptions about the coronal structure. This technique reconstructs the 3D electron density of the global corona using observations from multiple viewpoints provided by multiple spacecraft or solar rotation. Tomographic reconstructions of the 3D electron density have been demonstrated using white-light $pB$ images from the {\it Solar and Heliospheric Observatory} (SOHO)/{\it Large Angle and Spectrometric Coronagraph} (LASCO)-C2 \citep{fra02,fra07,fra10,vas08}, the {\it Solar Terrestrial Relations Observatory} (STEREO)/{\it inner coronagraph} (COR1) \citep{kra09,kra14,kra16a}, and the {\it Mauna Loa Solar Observatory} (MLSO)/{\it Mark-IV} (Mk4) \citep{but05,vas08}. By combining tomography techniques with spectroscopy of multithermal filters, a novel diagnostic tool called Differential Emission Measure Tomography (DEMT) was developed \citep{fra05, fra09}. DEMT has been used to reconstruct 3D electron density and temperature distributions in the corona by utilizing multiband extreme-ultraviolet (EUV) images from STEREO/{\it Extreme UltraViolet Imager} (EUVI) \citep{fra09,vas09,vas10,vas11,mac22} and the {\it Solar Dynamics Observatory} (SDO)/{\it Atmospheric Imaging Assembly} (AIA) \citep{nue15,kra16b,cho20}.  Comprehensive reviews on rotational tomography for 3D reconstructions of the white-light and EUV corona are provided by \citet{fra00,fra05,vas16}.

Reconstructing the 3D electron density of the solar corona using tomography involves discretizing the spatial density distribution function with numerical methods. This transforms the inversion of the density integral into a multidimensional linear problem. The method is similar to least squares, where the goal is to minimize the loss function between multi-directional observational data and the model's prediction. In tomography, the problem is often underdetermined, meaning there are more unknowns than knowns, so a regularization term is added to the cost function to ensure a unique and stable solution. 

For example, \citet{fra02} performed tomography inversions on a cylindrical grid, using a regularization term with second-order smoothing in the radial and axial directions, and first-order smoothing in the azimuthal angle, with three free regularization parameters. \citet{fra07} developed reconstructions on a spherical grid with only one regularization parameter, applying second-order smoothing in longitude and latitude, without radial smoothing. \citet{kra09} carried out tomography inversion on a Cartesian grid with first-order smoothing.

In this study, following the tomography method by \citet{kra09}, we develop inversion codes for both Cartesian and spherical grids with second-order smoothing, aiming to build a 3D electron density database over Solar Cycle 24 using STEREO/COR1 observations. Including radial weighting in the regularization term helps overcome oversmoothing at lower heights and prevents oscillatory artifacts at higher ones \citep{wan21,jon22}\footnote{\href{https://aas238-aas.ipostersessions.com/Default.aspx?s=7F-55-48-36-74-71-67-08-C5-5D-57-6E-D7-77-C7-5D}{238th AAS Meeting, iposter 328.07}}. Various tests show that this new approach improves the accuracy and quality of 3D global coronal density reconstructions. The mathematical approach, data preparation, grid usage, and results are detailed in subsequent sections.


\section{Method}
\label{sct:met}

\subsection{Regularized Tomography}
\label{ss:rtm}
The K corona forms due to Thomson scattering of photospheric white light from free electrons in the solar corona. Its polarized brightness ($pB$) can be expressed as an integration of electron density along the line of sight,
\begin{equation}
	pB(\rho)=\int_{\rm LOS} K(\rr)N_e(\rr)\dl,
	\label{eq:pb}
\end{equation}
where $N_e(\rr)$ is the electron density at a position $\rr$, and $\rho$ is the perpendicular distance between the LOS and Sun center. $K(\rr)$ is a known function related to the Thomson scattering effect \citep[e.g.,][]{bill66,que02,wan14}. 

By discretizing $N_e(\rr)$ using a linear combination of unknown densities $x_j$ for all grid points in the domain with index $j=1,...,n$, Equation~\ref{eq:pb} can be represented by the following linear equations \citep[e.g.,][]{fra00,fra02,kra09},
\begin{equation}
	\avec\xx = \yy,
        \label{eq:mod}
\end{equation}
where $\xx$ and $\yy$ are the column vectors. The element $y_i$ ($i=1,...,m$) of $\yy$ is the data value for the $i$-th ray in all used images.  The $n$ and $m$ represent the total number of elements in $\xx$ and $\yy$, respectively. The matrix $\avec$ contains the coefficients $a_{i,j}$ that are related to the integral of $K(\rr)$ along each ray.  

The practical method for solving Equation~\ref{eq:mod} involves using ridge regression, also known as Tikhonov regularization \citep{tik63}. This approach helps stabilize the solution and reduce artifacts caused by data noise and gaps. The regularized problem is addressed through minimization,
\begin{equation}
	\min_{\mathrm{x}}  F=\norm{\avec\xx-\yy}^2+\mu\norm{\rvec\xx}^2,
	\label{eq:min}
\end{equation}
where the matrix $\rvec$ represents a finite difference approximation, and $\mu$ is the regularization parameter, controlling the trade-off between the fidelity to the data and the regularization term. Increasing $\mu$ leads to a smoother solution, while decreasing $\mu$ allows the solution to fit the data more closely. We adopt the second-order smoothness using the second-order centered difference approximation,
\begin{equation}
	\norm{\rvec\xx}^2=\sum_{\ip,\jp,\kp}\left(\ppdr{f}{u}\right)^2_{\ip,\jp,\kp}+\left(\ppdr{f}{v}\right)^2_{\ip,\jp,\kp}+\left(\ppdr{f}{w}\right)^2_{\ip,\jp,\kp},
	\label{eq:reg}
\end{equation}	
where
\begin{eqnarray}
	\left(\ppdr{f}{u}\right)_{\ip,\jp,\kp} &=& \frac{f(\ip+1,\jp,\kp) -2f(\ip,\jp,\kp) +f(\ip-1,\jp,\kp)}{\Delta{u}^2},\\
	\left(\ppdr{f}{v}\right)_{\ip,\jp,\kp} &=& \frac{f(\ip,\jp+1,\kp) -2f(\ip,\jp,\kp) +f(\ip,\jp-1,\kp)}{\Delta{v}^2},\\
	\left(\ppdr{f}{w}\right)_{\ip,\jp,\kp} &=& \frac{f(\ip,\jp,\kp+1) -2f(\ip,\jp,\kp) +f(\ip,\jp,\kp-1)}{\Delta{w}^2}.
        \label{eq:cdf}
\end{eqnarray}
Here $f_{\ip,\jp,\kp}$ represents the density at a grid point $(\ip,\jp,\kp)$ in three-dimensional Cartesian coordinate system where $(u,v,w)=(x,y,z)$ or Carrington spherical coordinate system where $(u,v,w)=(r, \theta, \phi)$, and the $r$, $\theta$, and $\phi$ represent the radial distance, latitude, and longitude, respectively.  The $\Delta{u}$, $\Delta{v}$, and $\Delta{w}$ represent the grid scales in the three respective directions. In the finite difference calculation, we take $\Delta{u}=\Delta{v}=\Delta{w}=1$.

The simplest regularization is the zeroth-order smoothness by taking $\rvec={\vec I}$, an identity matrix, so the regularized problem becomes,
\begin{equation}
        \min_{\mathrm{x}}  F=\norm{\avec\xx-\yy}^2+\mu\norm{\xx}^2.
        \label{eq:reg0}
\end{equation}

The minimization problem described in Equation~\ref{eq:min} is equivalent to minimize a quadratic function below,
\begin{equation}
   F=\frac{1}{2}\xx^T\mvec\xx - {\vec b}^T\xx,
	\label{eq:quad}
\end{equation}
where $\mvec=\avec^T\avec+\mu\rvec^T\rvec$ and ${\vec b}=\avec^T\yy$. The solution $\xx$ can be determined from $\grad{F}=0$, i.e.,
\begin{equation}
	\mvec\xx = {\vec b},
	\label{eq:mxb}
\end{equation}
Given that $\mvec$ is an $n\times{n}$ symmetric positive definite (SPD) matrix, Equation~\ref{eq:mxb} has a unique solution. We solve Equation~\ref{eq:mxb} using the conjugate gradient (CG) method \citep{press93}. This approach is well-suited for solving large SPD systems. The CG method iteratively minimizes the quadratic function of Equation~\ref{eq:quad}, which naturally arises in least-squares problems. This allows us to efficiently compute $\xx$ without forming $\mvec^{-1}$ explicitly while leveraging the sparsity of $\avec$.

Since the standard CG method is unable to guarantee a density solution devoid of unphysical, negative values, we mitigate the occurrence of negative values in the reconstruction by selecting an appropriate regularization parameter through cross-validation (refer to Section~\ref{ss:cv}). We set the negative density values to zero in the final solution.

\subsection{Improvement by Radial Weighting}
\label{ss: irw}
\citet{kra09} found that a rapid decrease of pB signals with height may result in linear artifacts in the reconstruction and suggested to reduce this effect by applying a weighting factor in  the first term of Equation~\ref{eq:min} as 
\begin{equation}
	F^w_1=\sum_{i=1}^m\left[\left(\sum_{j=1}^n{w_ia_{ij}x_j}\right)-w_iy_i \right]^2,
	\label{eq:wf1}
\end{equation}	
where $w_i=1/I_{\rm bg}(r_i)$ is the weighting function and $I_{\rm bg}(r_i)$ the background pB at radial distance $r_i$. Here we determine $I_{\rm bg}(r)$ using a similar method as in \citet{kra09}: first for each image we extract the intensity profile $y(r,\phi)$ along a circle with radius of $r$, where $\phi$ is position angles. Then doing the Fast Fourier Transform (FFT) in $\phi$ to derive its Fourier spectrum $\hat{y}(r,k_\phi)$. Keeping the harmonics up to second order ($k_\phi=0,1,2$) and making the inverse FFT, we obtain $y^{FT2}(r,\phi)$ and its maximum amplitude $y_m^{FT2}(r)$. By averaging this amplitude for all images used in the reconstruction, we finally obtain the radial background pB $I_{\rm bg}(r)$ and use it in Equation~\ref{eq:wf1} for our density reconstructions (as default applied in all cases given in this study). An investigation of over 100 CRs from 2007 to 2014 indicates that the distribution of $I_{\rm bg}(r)$ derived from $y_m^{FT2}(r)$ closely matches that obtained through a simple average of intensity over the entire $\phi$ for all images used in each reconstruction during the rising and maximum phases of the solar cycle. However, during solar minimum, $I_{\rm bg}(r)$ primarily captures the distribution of streamers (belt).

\begin{SCfigure}
  \centering
           \includegraphics[width=0.5\textwidth,clip=]{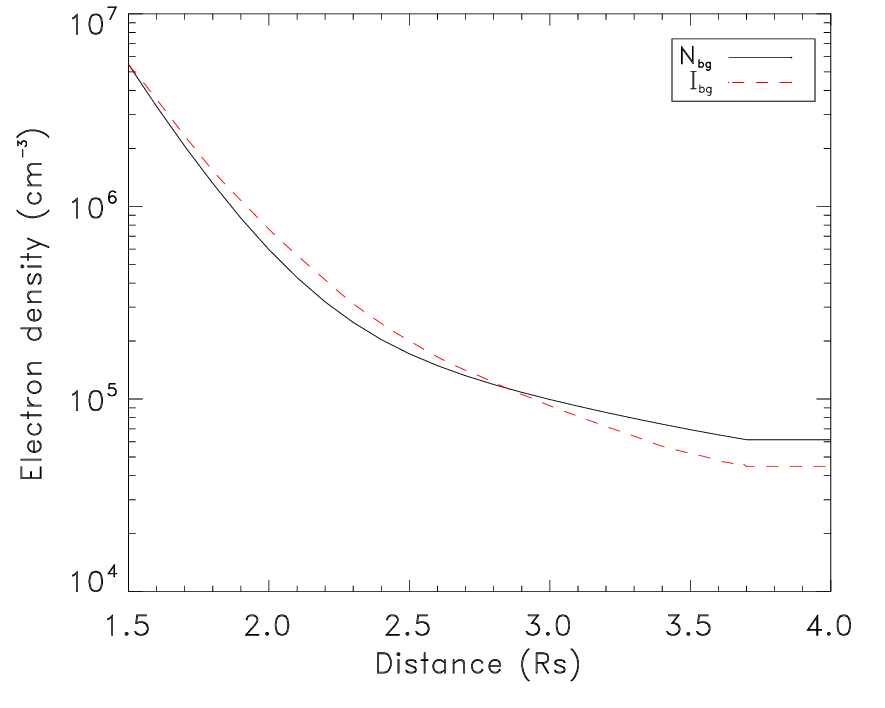}
	   \caption{Globally averaged electron density of the corona, $N_{bg}(r)$, as a function of radial distance $r$, within the range $1.5\leq(r/R_\odot)\leq 3.7$. The solid line represents $N_{bg}(r)$, obtained using the SSPA inversion method applied to STEREO COR1-B observations for CR 2098. The dashed line depicts the radial profile of background pB, $I_{bg}(r)$, normalized to the peak value of $N_{bg}(r)$. For radial distances beyond 3.7\rsun~(up to 4.0\rsun), both $N_{bg}(r)$ and $I_{bg}(r)$ are maintained at their respective values at r=3.7 \rsun.}
           \label{fig:rwtprf}
  \end{SCfigure}

While taking this approach can help recover the low-density structure at higher heights, the high-density structure near the inner boundary of the reconstruction would be significantly oversmoothed. On the other hand, the regularization with small values of $\mu$ favors to the recovery of more fine structure near the inner boundary, but it brings out oscillatory artifacts at larger heights. Here we propose to overcome this defect by adding a weighting factor $w = 1/N_{bg}(r)$ or $1/I_{\rm bg}(r)$ in the regularization term as, 
\begin{align}
	F^w_2 &= \norm{\rvec^w\xx}^2 \notag \\
	  &= \sum_{\ip,\jp,\kp} w^2(r_{\ip,\jp,\kp}) \left[ \left(\ppdr{f}{u}\right)^2_{\ip,\jp,\kp} + \left(\ppdr{f}{v}\right)^2_{\ip,\jp,\kp} + \left(\ppdr{f}{w}\right)^2_{\ip,\jp,\kp} \right],
        \label{eq:wrg}
\end{align}
where $N_{bg}(r)$ is the radial profile of background electron density, calculated by global average of 3D density model obtained by the spherically symmetric polynomial approximation (SSPA) method \citep{wan17}. $I_{\rm bg}(r)$ is the radial background pB, same as used in the first term of minimization. Figure~\ref{fig:rwtprf} demonstrates the distributions of $N_{bg}(r)$ and $I_{\rm bg}(r)$ for CR 2098. We found that applying the radial density or pB weighting gives nearly identical results (see Section~\ref{sst:irw}). 

\subsection{Cross Validation}
\label{ss:cv}
  Cross-validation (CV) is a general method for selecting the value of regularization parameter in tomography reconstruction, and it is also widely used in mechine learning for training a model to predict data. Following applications of this method in reconstructions of the coronal density using pB images from LASCO/C2 \citep{fra02} and STEREO/COR1 \citep{kra09}, we propose a $n$-fold CV approach as described below: 

\begin{enumerate}
 \item Randomly extract 10$-$20\% from the rows of the matrix $\avec$ to make a sampling matrix $\avec_s$, and extract the corresponding elements from $\yy$ to make a sampling vector $\yy_s$. 
 \item  Invert the solution $\hat{\xx_i}$ from the remaining data ($\avec^\prime$ and $\yy^\prime$) for various choices of the regularization parameter $\mu_i$. 
 \item Repeat Steps (i) and (ii) $n$-times so obtaining $n$-fold solutions ($\hat\xx_i^1$, $\hat\xx_i^2$, ..., $\hat\xx_i^n$) for each $\mu_i$. Then estimate the mean error $\bar{\chi}_i$ for the model-predicted data by:
	 \begin{equation}
		 \bar{\chi_i}=\left[\frac{1}{n}\sum_{k=1}^n\norm{\avec_s^k\hat\xx_i^k-\yy_s^k}^2\right]^{1/2}.
	 \end{equation}
\end{enumerate}
We determine the optimal value of the regularization parameter, denoted as $\mu_{\rm best}$, by identifying the one that yields the smallest $\bar{\chi}$. The standard deviations, computed for the average of $n$-fold solutions using $\mu_{\rm best}$, serve as indicators of the uncertainty in the reconstructed density. Tests reveal that the reconstructed density with $\mu_{\rm best}$ occasionally includes some small negative values (approximately a few percent, mostly near zero), which diminish as $\mu$ increases. Since all elements in $\avec$ are positive, converting negative elements in $\xx_i^k$ to positive can enhance their contributions to $\bar{\chi_i}$, aiding in the determination of $\mu_{\rm best}$ associated with solutions exhibiting fewer negative values. We take this strategy in this study. Previous studies have suggested that such negative-density artifacts (or zero-density regions when applying an algorithm with the non-negative solution) may occur due to fast changes of the density \citep{fra05} or very low true density there (smaller or comparable to data noises such as in coronal holes) \citep{kra09}. 

\begin{figure}
  \centerline{\includegraphics[width=1.0\textwidth,clip=]{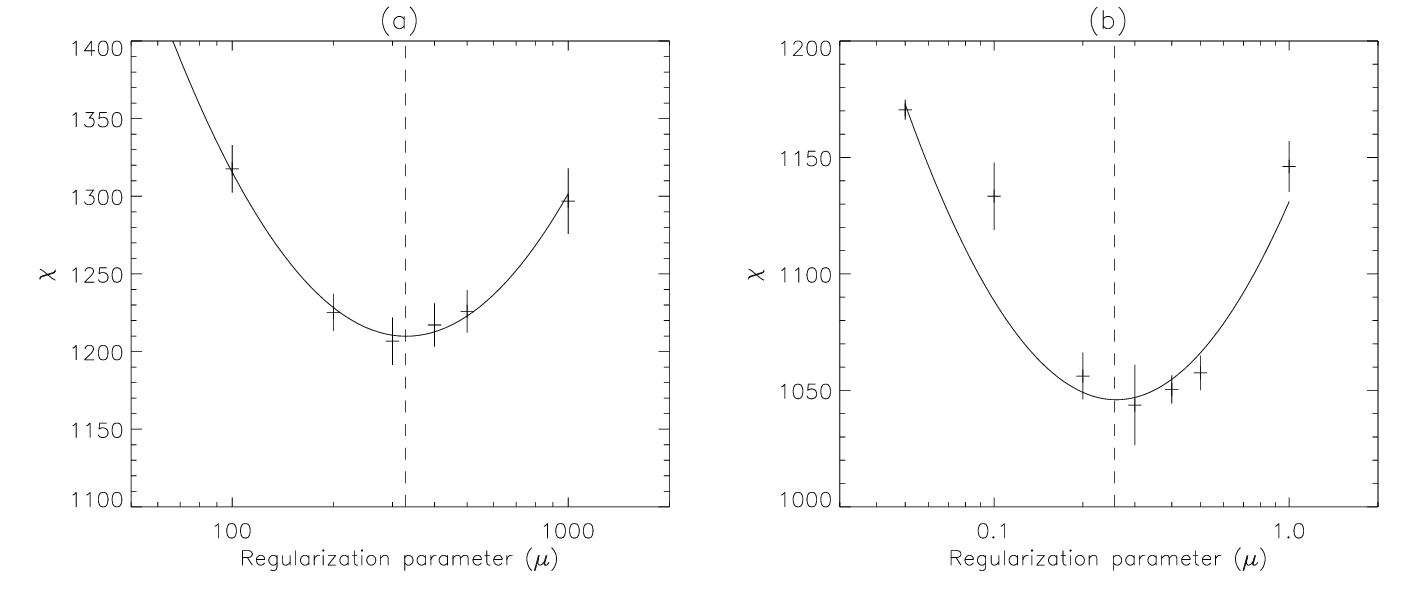}}
  \caption{Determination of the optimal regularization parameter by fitting a parabola to $\chi$ for solutions with varying $\mu$ values. (a) Reconstruction of CR 2098 without radial weighting in the regularization term. The vertical dashed line indicates $\mu_{\rm best}=330$. (b) Reconstruction with radial weighting in the regularization. The vertical dashed line indicates $\mu_{\rm best}=0.26$. Error bars in (a) represent the standard deviations of the $\chi$ values from the 7-fold cross-validation average, while those in (b) are derived from the 5-fold cross-validation average.}
          \label{fig:mfit}
  \end{figure}

Figure~\ref{fig:mfit} illustrates the determination of $\mu_{\rm best}$ by fitting $\chi$ as a function of $Log_{10}(\mu)$ to a parabolic curve to find the minimum point. The values of $\bar{\chi_i}$ were calculated as the average of 7-fold solutions (panel a) or 5-fold solutions (panel b) with a data sampling rate of 20\%.

\section{Data Preparation}
\label{sct:dat}
The STEREO, a pair of identical satellites, was launched on 25 October 2006, with each spacecraft orbiting the Sun in opposite directions to provide a stereoscopic view of the Sun \citep{how08}. The STEREO-A spacecraft passed by Earth again on 12 August 2023 for the first time, while the STEREO-B spacecraft lost communication on 1 October 2014 and eventually became non-responsive. COR1 is a white-light coronagraph that observes the inner corona from 1.3 to 4 solar radii, forming part of the {\it Sun Earth Connection Coronal and Heliospheric Investigation} (SECCHI) instrument suite \citep{thom03,thom08} onboard the STEREO spacecraft. The STEREO/COR1 coronagraphs play a critical role in advancing our understanding of solar eruptions, solar wind, and space weather forecasting through detailed imaging of the solar corona and stereoscopic observations of CMEs.

In this study, we reconstruct 3D coronal densities using pB observations from STEREO/COR1-B. Since the white-light emission of the K-corona is optically thin, reconstructing the 3D coronal density requires data collection over half a solar rotation when observed from a single satellite. We sample images at a 12-hour cadence from the COR1-B database. Images with CME activity or significant instrument noise (such as ring-shaped and cloud-like artifacts) are replaced with clean, relatively stable images taken within $\pm$12 hours, identified through visual inspection. To stabilize the inversion process for the chosen grid resolution (see Section~\ref{sct:inv}), we maintain a minimum separation of about $2^\circ$ (corresponding to a cadence of 3.6 hours) between the observing directions of consecutive images. Based on this criterion, approximately 27$-$28 pB images are typically selected for reconstruction.

Before 19 April 2009, the images were recorded in a 1024$\times$1024, while after that date, they were recorded in a 512$\times$512 format. The instrumental scattered light in the pB data is removed by subtracting the combined monthly minimum and calibration roll backgrounds \citep{thom10,wan17}. Following despike processing to remove cosmic rays and ``hot pixels" using the SSW routine {\scshape ssw\_unspike\_cube}, the images are rebinned into a 128$\times$128 format. The unit vector of the LOS direction associated with each pixel (in Carrington longitude and latitude) is determined by transforming its position vector relative to the observer from the image coordinate system into the Carrington coordinate system \citep[see][]{kra06}. The background pB ($I_{\rm bg}(r)$) is calculated by averaging all 128$\times$128 images used in the tomographic reconstruction. 

\begin{SCfigure}
  \centering
  \includegraphics[width=0.6\textwidth]{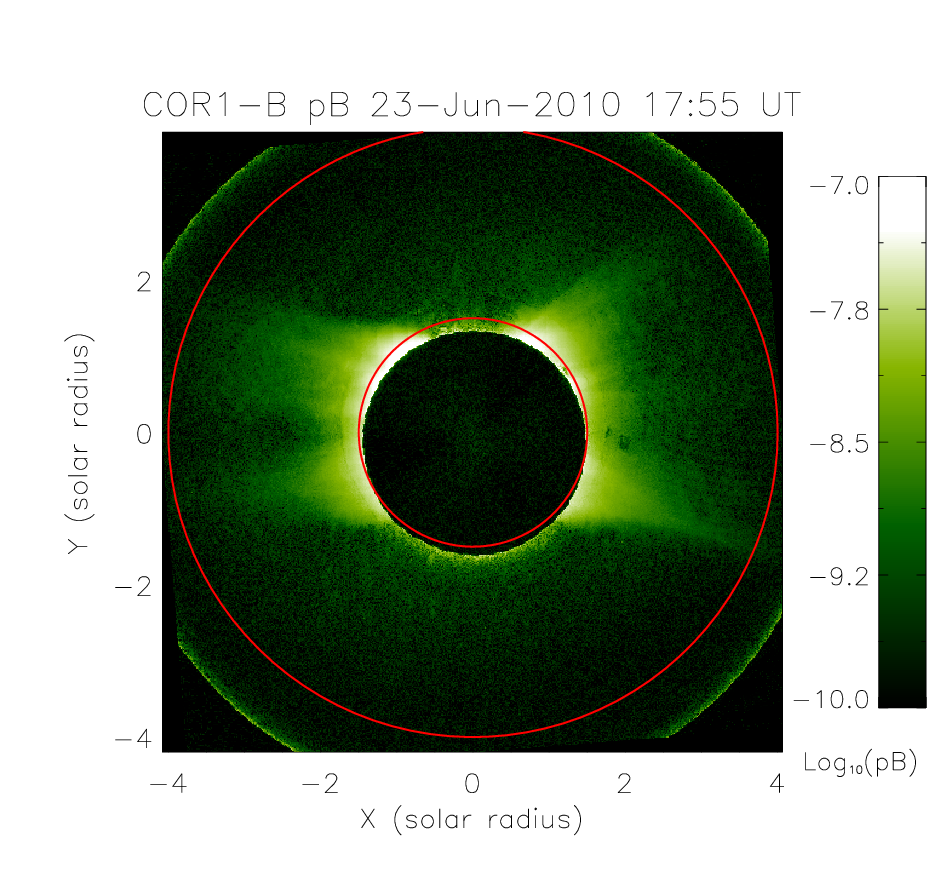}
	  \caption{ Illustration of a polarized brightness (pB) image used for the density reconstruction of CR 2098, processed with despike correction. Two red circles mark the inner (1.5\rsun) and outer (4.0\rsun) boundaries of the region from which input data are sampled. An animation of all the pB images used is available in the online journal.}
          \label{fig:pbmap}
  \end{SCfigure}

In the following analysis, we demonstrate the 3D density reconstruction by tomography using 28 pB images observed from 23 June 2010, 17:55 UT to 7 July 2010, 00:05 UT (covering the middle of CR 2098) with STEREO/COR1-B (see Figure~\ref{fig:pbmap}). We assess this density reconstruction by comparing it with a global thermodynamic MHD model provided by Predictive Science Inc. \citep[PSI;][]{mik99,lion09}. Additionally, we demonstrate two consecutive reconstructions for CR 2091, from 7 December 2009 to 3 January 2010 (see Figure~\ref{fig:n3view}). Each half-CR reconstruction is made using 28 pB images from COR1-B.

\section{Inversion on Cartesian and Spherical Grids}
\label{sct:inv}

The inversion of 3D electron density can be implemented using various types of grids such as Cartesian \citep{kra09}, spherical  \citep{fra00}, or cylindrical grids \citep{fra02}. In this study we demonstrate the first two. 

In the Cartesian case, we use a $128^3$ grid in Carrington coordinates with a spherical domain of $R_{\rm in}-2ds \le {r}\le R_{\rm out}+2ds$, where $R_{\rm in}=1.5 R_\odot$ and $R_{\rm out}=4.0 R_\odot$ are the inner and outer boundaries of data points in each image, and $ds=0.063 R_\odot$ is the grid size. We choose the radial range for the solution domain slightly larger than the data range to reduce boundary effects. The total number of data points (rays) sampled in a region of $R_{\rm in} \le {r}\le R_{\rm out}$ for 28 rebinned $128\times128$ images is $m=295,549$ for the reconstruction of CR 2098. This results in a sparse matrix $\avec$, where the number of nonzero elements is $3.97\times10^{-4}$ times the total number of elements. In the solution domain, there are  $n=1,124,072$ density elements (voxels), giving a data point to density element ratio of $m/n=26.3\%$. Thus, Equation~\ref{eq:mod} is an underdetermined problem, and adding a regularization term is essential to ensure a unique solution. Additionally, keeping the separation between the observing directions of consecutive images larger than the angular size of voxels relative to the Sun's center helps stabilize the inversion. Our selected images with a typical 12-hour sampling cadence meet this condition well. 

In the spherical grid case, we use a grid of $361\times181\times51$ voxels in the longitudinal, latitudinal, and radial directions. This gives a bin size of $1^{\circ}$ in longitude and latitude, and a radial bin size of 0.05 $R_\odot$. The inversion occurs within the domain of $R_{\rm in} \le {r}\le R_{\rm out}$. The data sampling is the same as in the Cartesian grid case, with a data point to density element ratio estimated to be $m/n=8.9\%$. In the reconstructions of CR 2091, the resulting sparse matrix $\avec$ contains nonzero elements amounting to $2.19\times10^{-4}$ of the total number of elements. Tests using a reduced grid resolution, $181\times91\times51$, but the same image sampling cadence, and using the same grid ($361\times181\times51$) but a higher sampling cadence of 6 hours, produce similar solutions for $\mu_{\rm best}$. This indicates that a $181\times91\times51$ grid with a 12-hour image cadence is sufficient for density reconstruction using observations from a single satellite. Increasing the image cadence does not improve the spatial resolution of reconstructed density, suggesting that smaller density structures may be smoothed out due to evolution during the data accumulating period \citep[see][]{fra07}.

For the relatively small field-of-view (FOV) of COR1 observations, density reconstructions using Cartesian and spherical grids show nearly no difference. Tomographic reconstruction in a Cartesian grid (e.g., for ray tracing) is much simpler than in a spherical grid. However, when using observations over a larger FOV, it requires significantly more computational resources compared to a spherical grid, especially when a uniform grid algorithm is employed.

\begin{figure}
  \centerline{\includegraphics[width=1.0\textwidth,clip=]{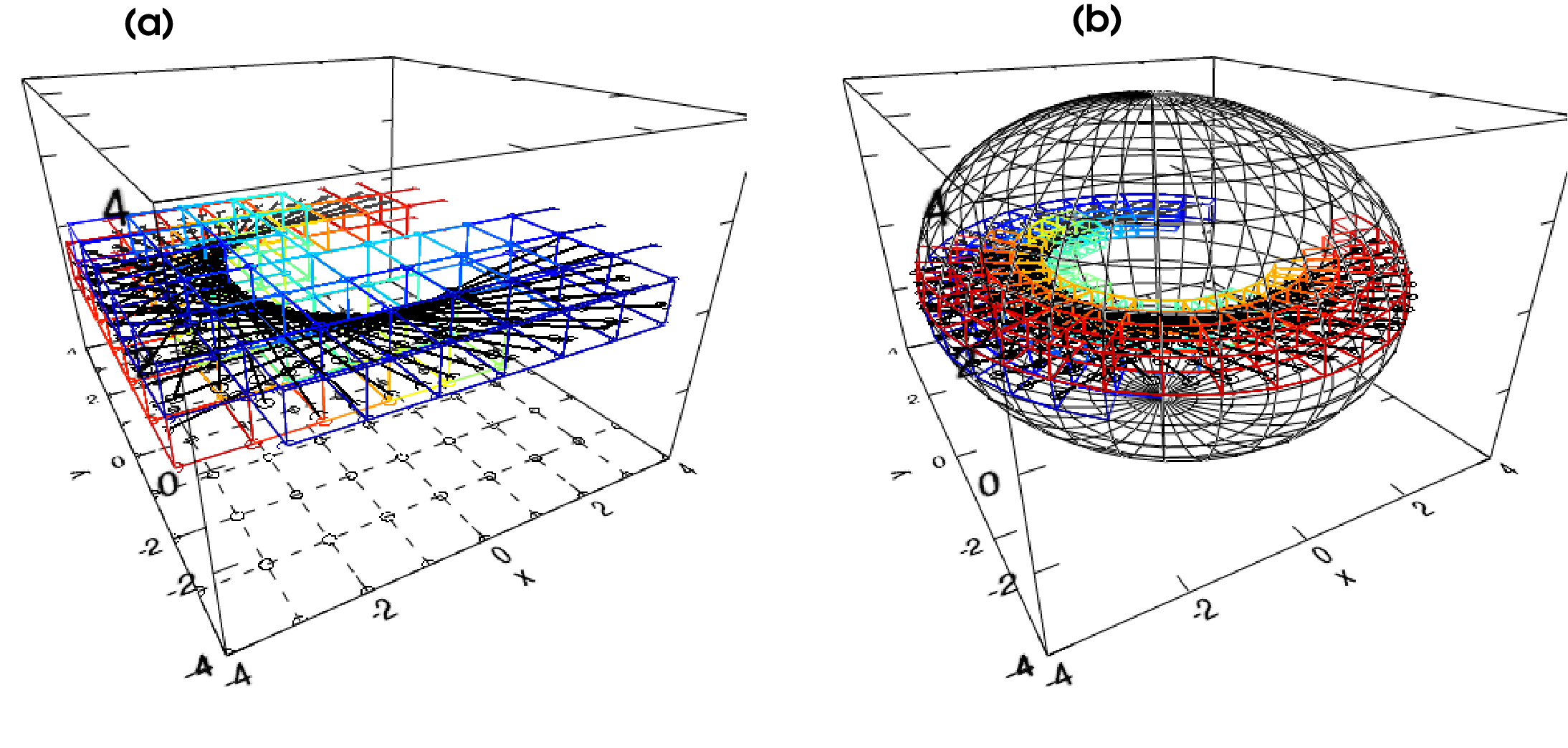}}

  \caption{Illustration of the tomographic reconstruction of coronal density in (a) Cartesian grid and (b) spherical grid.
Panel (a): Displays a $9^3$ Cartesian grid with coordinates $x$, $y$, and $z$ ranging from $-4$ to $4 \,R_\odot$. The 28 rays, corresponding to a given data point from 28 successively observed images over two weeks by COR1-B, are shown as black lines, while the intersected cells are indicated by colored lines.  Panel (b): Depicts a grid of $37\times19\times6$ in longitude, latitude, and radial direction ($1.5 \leq r \leq 4.0 \,R_\odot$). The black lines represent the same 28 rays as in (a), with the intersected cells similarly highlighted in color.} 
          \label{fig:grid}
  \end{figure}

Figure~\ref{fig:grid} demonstrates the 3D coronal electron density reconstruction from 28 successively observed pB images over two weeks using a $9^3$ Cartesian grid (panel a) and a $37\times19\times6$ spherical grid (panel b). 

\section{Results}
\label{sct:res}

\subsection{Effects of Regularization Parameters}
\label{sst:erp}
\begin{figure} 
  \centerline{\includegraphics[width=1.0\textwidth,clip=]{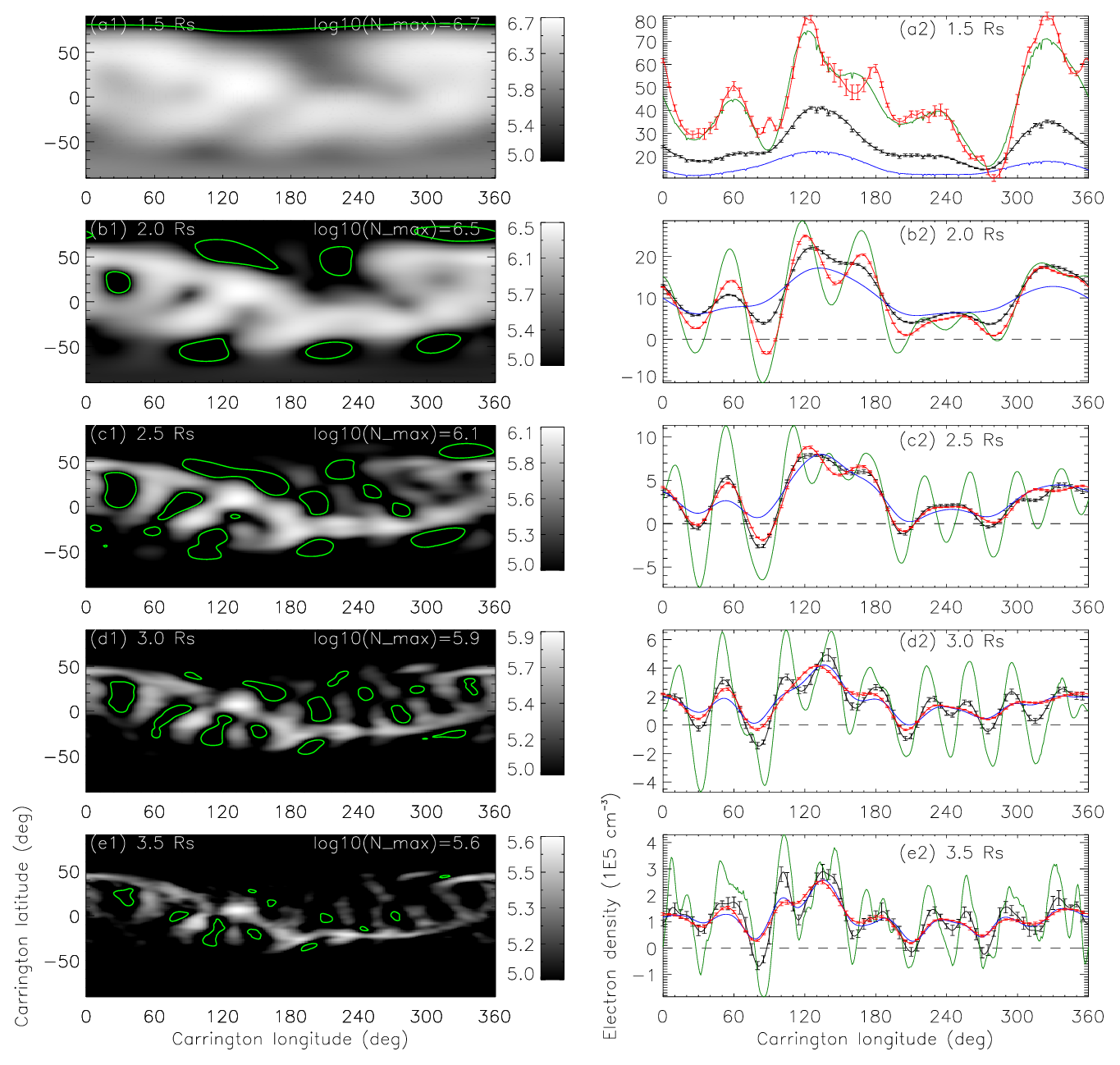}} 
	  \caption{Reconstruction of the 3D coronal electron density for CR 2098 using tomography with a regularization parameter of $\mu=300$. {\it Left panels}: Spherical cross sections of the density in logarithmic scale at 1.5, 2.0, 2.5, 3.0, and 3.5 $R_\odot$. The contours indicate the regions of non-positive values. {\it Right panels}: Corresponding density profiles along the equator (black lines). Error bars are determined by cross-validation (see text for details). The green and blue lines represent reconstructions with $\mu=10$ and $\mu=4000$, respectively, while the red lines show the case with $\mu=0.3$ and radial density weighting applied.}
          \label{fig:nmap_mbt}
  \end{figure}

Figure~\ref{fig:nmap_mbt} shows the results of the 3D electron density reconstruction for CR 2098 using the tomography technique with  a $128^3$ grid and second-order smoothness (see Equations~\ref{eq:reg}$-$\ref{eq:cdf}). The regularization parameter $\mu=300$ was determined by 7-fold cross-validation (see Figure~\ref{fig:mfit}a). The left panels of Figure~\ref{fig:nmap_mbt} display spherical cross-sections of the electron density at different heliocentric distances ($r=1.5-3.5$ \rsun), interpolated from the solution in the Cartesian grid. The right panels show the density profiles along the equator, with error bars representing the standard deviation of the mean density calculated from the 7-fold solutions. For instance, for data points between longitudes $120^{\circ}-180^{\circ}$, the average relative errors are estimated to be 1.7\%, 1.8\%, 3.2\%, 10\%, and 13\% at 1.5, 2.0, 2.5, 3.0, and 3.5 \rsun, respectively. Negative values appear in some regions, as shown by the contours in the density maps at different heights (see Figure~\ref{fig:nmap_mbt}), which could result from the rapid evolution of small density structures \citep{fra02} or data with a lower signal-to-noise ratio (S/N). Tests show that the negative values in the solution can be mostly avoided by using a sufficiently large $\mu$ value, which will smooth out small density structures.   

\begin{figure}
  \centerline{\includegraphics[width=1.0\textwidth,clip=]{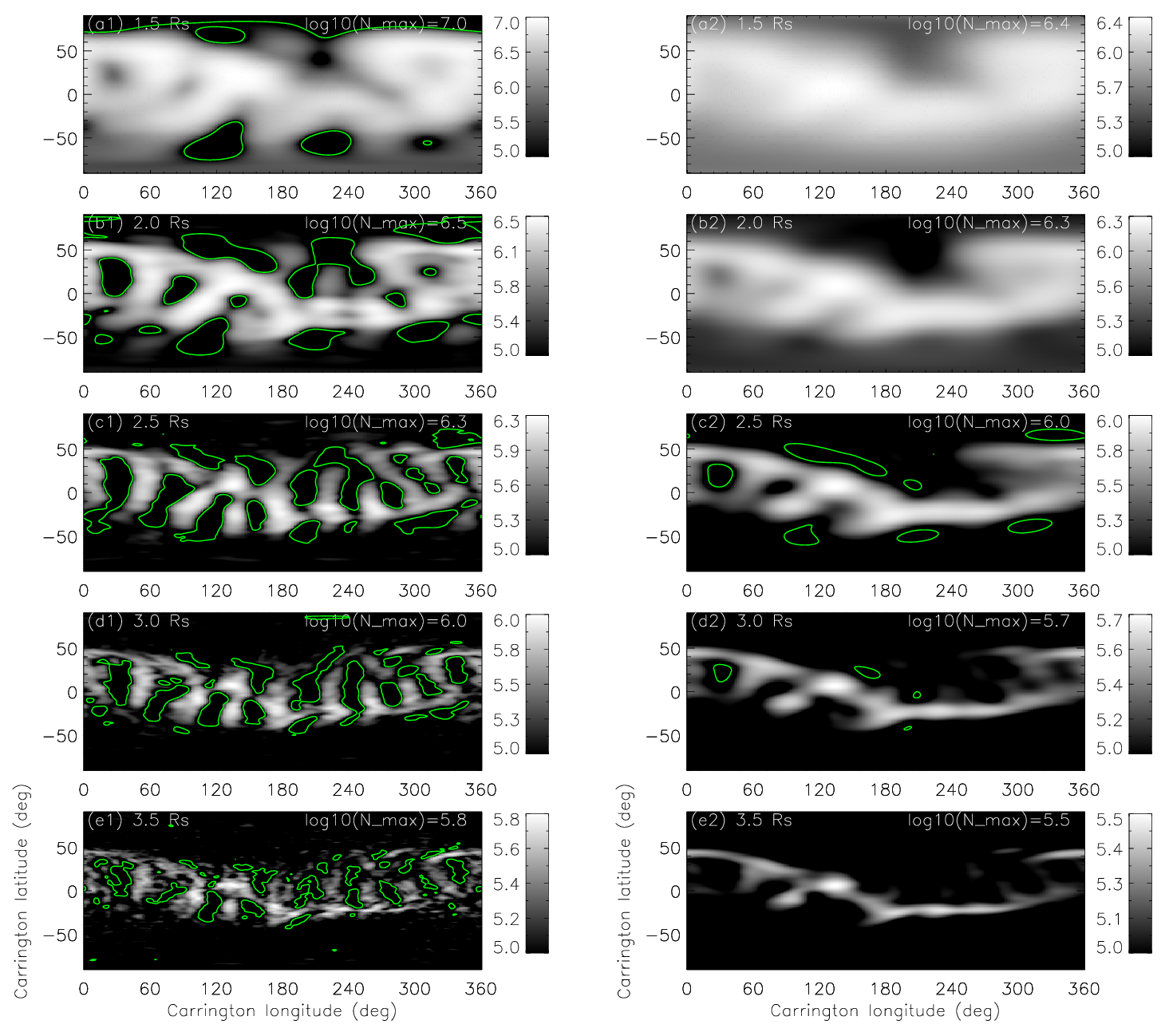}}
\caption{Comparison between density reconstructions with $\mu=10$ ({\it Left panels}) and $\mu=4000$ ({\it Right panels}) in the case without applying radial weighting. Panels from top to bottom display the density distributions in logarithmic scale at 1.5, 2.0, 2.5, 3.0, and 3.5 $R_\odot$. The contours indicate the regions of non-positive values. }
\label{fig:nmap_mcp}
  \end{figure}

The right panels of Figure~\ref{fig:nmap_mcp} show the density distributions at different heights for the case with $\mu$=4000. The corresponding density profiles along the equator are shown in blue in the right panels of Figure~\ref{fig:nmap_mbt}. This clearly indicates that the regions with negative values are significantly reduced for this solution compared to the case with $\mu=300$. However, this result is achieved at the expense of over-smoothing the larger density structure at lower heights (see panel a2 and b2). Conversely, choosing small $\mu$ values (e.g., $\mu$=10) can recover more fine structure at lower heights (see Figure~\ref{fig:nmap_mcp}a1 and b1), but it introduces oscillatory artifacts at higher heights (Figure~\ref{fig:nmap_mcp}c1-e1) and significantly increases the number and amplitudes of negative values (see the contours in Figure~\ref{fig:nmap_mcp}c1-e1 and the green curves in Figure~\ref{fig:nmap_mbt}c2-e2). 

We find that the quality of the density reconstruction can be significantly improved by incorporating a radial weighting factor in the regularization term (see Equation~\ref{eq:wrg}). This approach efficiently suppresses the occurrence of negative values at higher altitudes while also recovering the fine and large density structures at lower heights.

\subsection{Improved Reconstruction by Radially Weighted Regularization}
\label{sst:irw}
\begin{figure}
  \centerline{\includegraphics[width=1.0\textwidth,clip=]{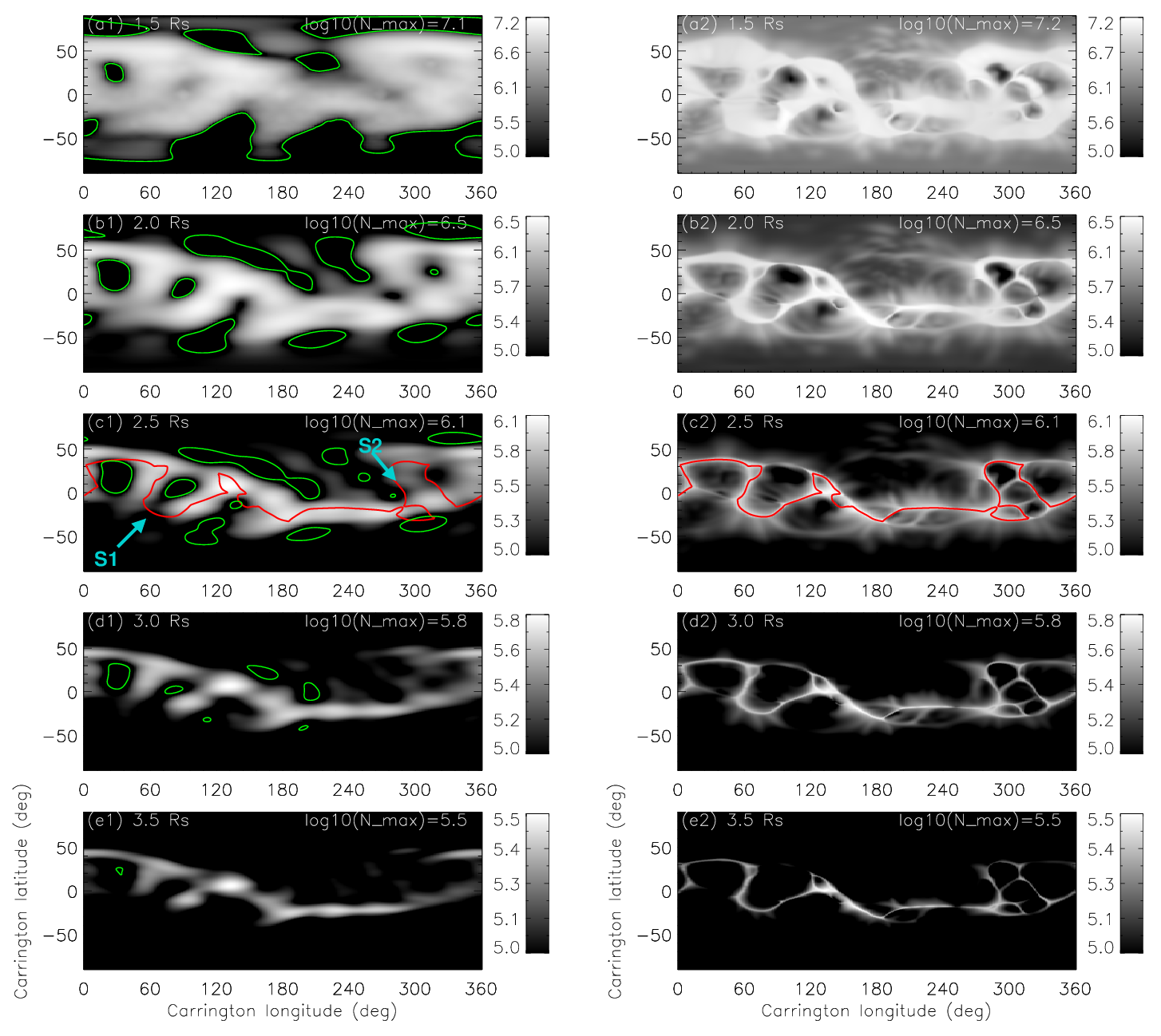}}
\caption{Comparison between the 3D coronal density for CR 2098 derived using radially-weighted tomography with $\mu=0.3$ ({\it Left panels}) and that for CR 2097/2098 from the MHD simulation ({\it Right panels}). Panels from top to bottom show the density distributions on a logarithmic scale at 1.5, 2.0, 2.5, 3.0, and 3.5 $R_\odot$. In the left panels, the green contours indicate the regions of non-positive values. In (c1) and (c2), the red contours represent the magnetic neutral line for the radial component of the magnetic fields in the MHD model. Arrows S1 and S2 in (c1) mark two segments of the magnetic neutral line discussed in detail in the text.}
            \label{fig:ntmd}
  \end{figure}

The left panels of Figure~\ref{fig:ntmd} show the results of the 3D coronal density reconstruction for CR 2098 using the improved tomography technique with a radial weighting factor $w=N_m/N_{bg}(r)$ in the regularization term, where $N_{bg}(r)$ is the radial background density calculated from the SSPA method (see Figure~\ref{fig:rwtprf}), and $N_m$ is the peak value. The regularization parameter $\mu=0.3$ was determined by 5-fold cross-validation (see Figure~\ref{fig:mfit}b). In panel a1, small, weak ``dot"-like enhanced features at the longitudes of 90$^{\circ}$, 180$^{\circ}$, and 270$^{\circ}$ along the equator are noted. These are inner boundary artifacts resulting from the imposition of spherical boundaries on the solution in the Cartesian grid. No such boundary artifacts are present in the solution obtained using the spherical grid (see Section~\ref{sst:rsg}). Compared to the reconstructions without radial weighting, the density distributions at $r$= 1.5 and 2.0 \rsun\, are similar to those for the reconstruction with $\mu$=10 (compare panel a1 and b1 between  Figures~\ref{fig:ntmd} and \ref{fig:nmap_mcp}), while the density distributions at $r$= 2.5, 3.0, and 3.5 \rsun\, are similar to those for $\mu$=4000 (compare Figure~\ref{fig:ntmd}c1-e1 with Figure~\ref{fig:nmap_mcp}c2-e2). Compared to the non-radial weighting reconstruction with the optimized regularization parameter ($\mu$=300), the radial-weighting reconstruction recovers more fine density structures near the inner boundary (compare panel a1 and b1 between  Figures~\ref{fig:ntmd} and \ref{fig:nmap_mbt}) while showing less influence by oscillatory artifacts and fewer negative-value voxels in the density distribution at higher heights (compare panel d1 and e1 between  Figures~\ref{fig:ntmd} and \ref{fig:nmap_mbt}).

The right panels of Figure~\ref{fig:nmap_mbt} show the equatorial density profiles at various heights for the reconstruction with radial weighting (red lines), with error bars indicating the standard deviation of the mean density calculated from the 5-fold solutions. For data points between longitudes $120^{\circ}-180^{\circ}$, the average relative errors are estimated to be 5.1\%, 1.4\%, 1.9\%, 2.6\%, and 4.2\% at 1.5, 2.0, 2.5, 3.0, and 3.5 \rsun, respectively. Notably, the relative errors at 3.0 and 3.5 \rsun~ decrease to below 5\%, compared to values exceeding 10\% in the case without radial weighting. This reduction is likely due to the effects of radial weighting, which enhances the smoothing of densities at greater heights. Conversely, the reduced smoothing at lower heights slightly increases the relative errors at 1.5 \rsun. In Figure~\ref{fig:nmap_mbt}a2, weak boundary artifacts appear as small bumps at longitudes 90$^{\circ}$ and 270$^{\circ}$. 

For comparison between the cases with and without radial weighting, the equatorial density profiles for the non-radial weighting solutions with $\mu$=10 (green lines), 300 (black lines), and 4000 (blue lines) are shown in the right panels of Figure~\ref{fig:nmap_mbt}. The density profile at 1.5\rsun\, in the radial weighting case is consistent with that of the non-radial weighting case with $\mu$=10 (with less smoothing). At 2.5\rsun, the density profiles for the radial and non-radial weighting case with $\mu$=300 are comparable. However, at greater heights of 3.0 and 3.5 \rsun, the radial weighting density profiles align more closely with those of the non-radial weighting case with $\mu$=4000 (greater smoothing) and show significantly fewer negative values (see also Figure~\ref{fig:ntmd}d1 and e1). The radial weighting clearly enhances the recovery of densities at 1.5\rsun, where the peak amplitudes are approximately double those in the non-radial weighting case with the optimal $\mu$=300. 

\begin{figure}
  \centerline{\includegraphics[width=1.0\textwidth,clip=]{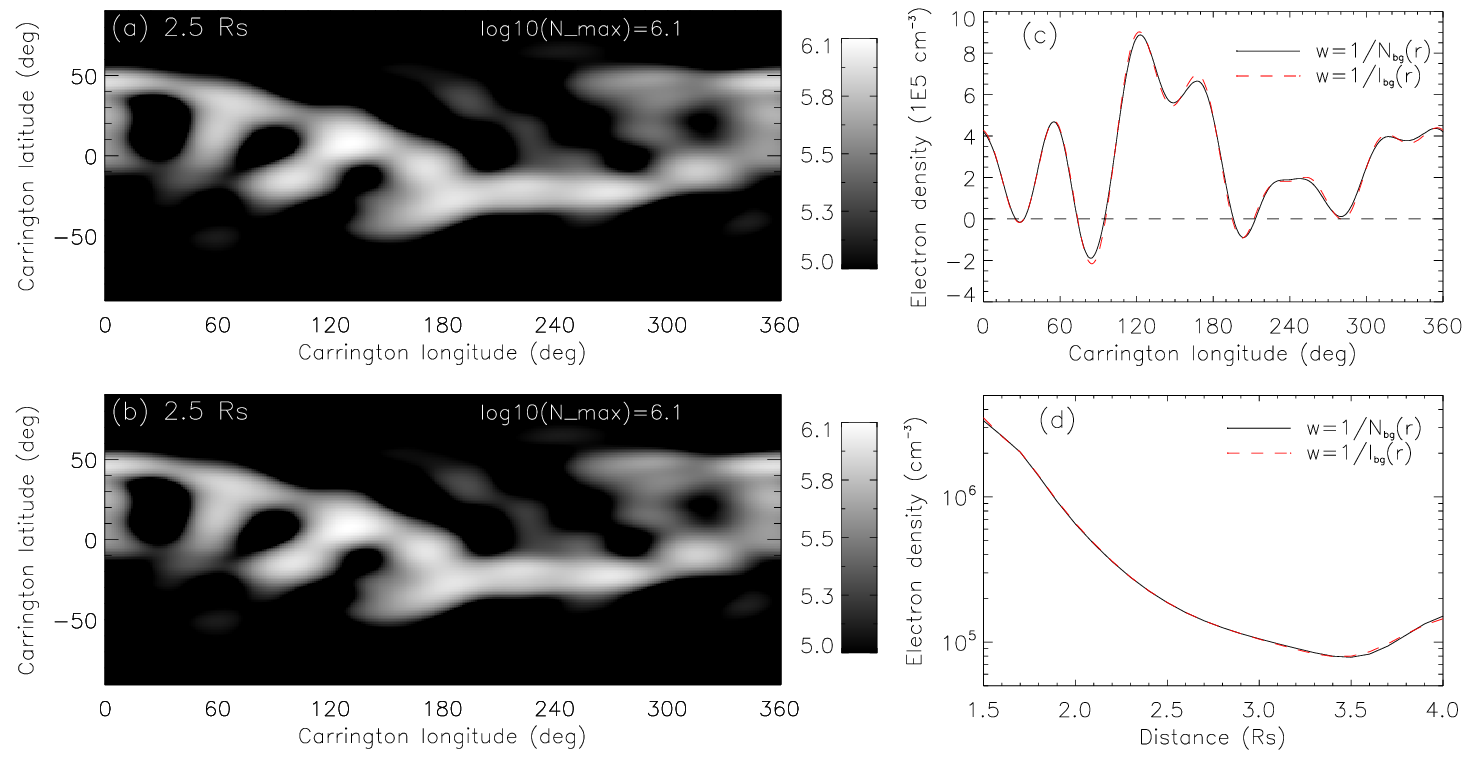}}
\caption{Comparison between 3D coronal density reconstructions obtained by tomography using the radial profile of background density as weights ($w=1/N_{bg}(r)$) and the radial profile of background pB as weights ($w=1/I_{bg}(r)$). (a) Density distribution at 2.5\,$R_\odot$ with density weighting. (b) Density distribution at 2.5\,$R_\odot$ with pB weighting. (c) Density profiles along the equator at 2.5\,$R_\odot$ for both cases. (d) Radial profiles of globally averaged density for both cases. In (c) and (d), the black solid line represents the density weighting case, while the red dashed line represents the pB weighting case. Note that in (c) and (d), the two curves are nearly overlaid.}
            \label{fig:npbwt}
 \end{figure}

Figure~\ref{fig:npbwt} compares the density reconstructions between the cases using radial density weighting and pB intensity weighting with the same regularization parameter $\mu=0.3$. The radial intensity weighting factor $w=I_m/I_{bg}(r)$ was used in regularization, where $I_m$ is the peak background intensity. The quantitative comparisons of density distributions at different heights (e.g., the case at $r=2.5$\rsun~ shown in panel a and b), density profiles along the equator (panel c), and globally averaged radial density profiles (panel d) show that the reconstructions in the two cases are nearly identical. This indicates that the reconstructed solution is insensitive to small differences between the normalized weighting functions $N_{bg}(r)$ and $I_{bg}(r)$, suggesting that we can choose weighting functions used in the two terms of Equation~\ref{eq:min} to be the same as $w=I_m/I_{bg}(r)$ for convenience. This is because $I_{bg}(r)$ can be directly obtained from the data, while the calculation of $N_{bg}(r)$ also depends on the model assumption \citep{wan17}.      

In addition, calculating $N_{bg}(r)$ directly from the tomographic reconstruction without a weighting factor could provide a self-consistent approach. However, beyond the concern of optimizing computational time, another issue arises from an artificial effect: a tendency for density to increase near the outer boundary due to the assumption of a finite computational domain (see Figure~\ref{fig:npbwt}d and Figure~\ref{fig:nrcmp}). This limitation, however, can be mitigated by using the SSPA method or $I_{bg}(r)$ obtained from pB images.

\subsection{Comparison with MHD Density Model}
 \label{sst:cmm}
In this section, we evaluate the density reconstruction for CR 2098 obtained through improved tomography using the Corona Heliosphere (CORHEL) and Magnetohydrodynamics Around a Sphere (MAS) model, known as the CORHEL MAS model, developed by PSI \citep[see][]{mik07,lion09}. The global plasma density and temperature structures modeled using the MAS code have been successfully applied to predict white-light coronal features for various total solar eclipses \citep[e.g.,][]{rus10,mik18}. Specifically, the model was used to predict the coronal structure of the 11 July 2010 eclipse, incorporating photospheric magnetic field data from SOHO/MDI collected between 10 June and 4 July 2010 (spanning CR 2097 and 2098) as boundary conditions. The resulting 3D coronal density model was also validated against the density reconstruction of CR 2098 derived from STEREO/COR1-A and -B observations using the SSPA method \citep{wan17}.

Figure~\ref{fig:ntmd} compares the density distributions at different heights between the tomography-reconstructed (left panels) and MHD-modeled (right panels) 3D densities for CR 2098, both displayed on the same logarithmic scale for each height. We find that the tomographic density structures are roughly consistent with those predicted by the MHD model. However, some minor differences are still clearly visible. For example, the density structures in the MHD model exhibit more fine features compared to those in the tomographic model, and the main high-density structures (e.g., those along the magnetic neutral line in the MHD model) are distinctly narrower. Considering the tomographic reconstruction was based on data observed from a single satellite over a period of two weeks, the presence of these minor differences are reasonable.  It is evident from panel c2 that the highest density structures tend to follow the magnetic neutral line. By taking advantage of this characteristic, \citet{jon22} developed a technique to constrain global coronal models based on the 3D tomographic density reconstructions. Assuming that the density structures recovered by tomography are those that can survive over two weeks, we argue that the MHD model may not predict the accurate positions of the magnetic neutral line in two segments (marked by arrows S1 and S2 in panel c1), because no high density features are observed following the magnetic neutral line at these two regions. This discrepancy could be attributed to the MHD model's reliance on a static boundary condition, using photospheric magnetic field measurements collected over approximately a month. This duration is much longer than the data observation period required for tomography reconstructions.  

\begin{SCfigure}
  \centering
            \includegraphics[width=0.6\textwidth,clip=]{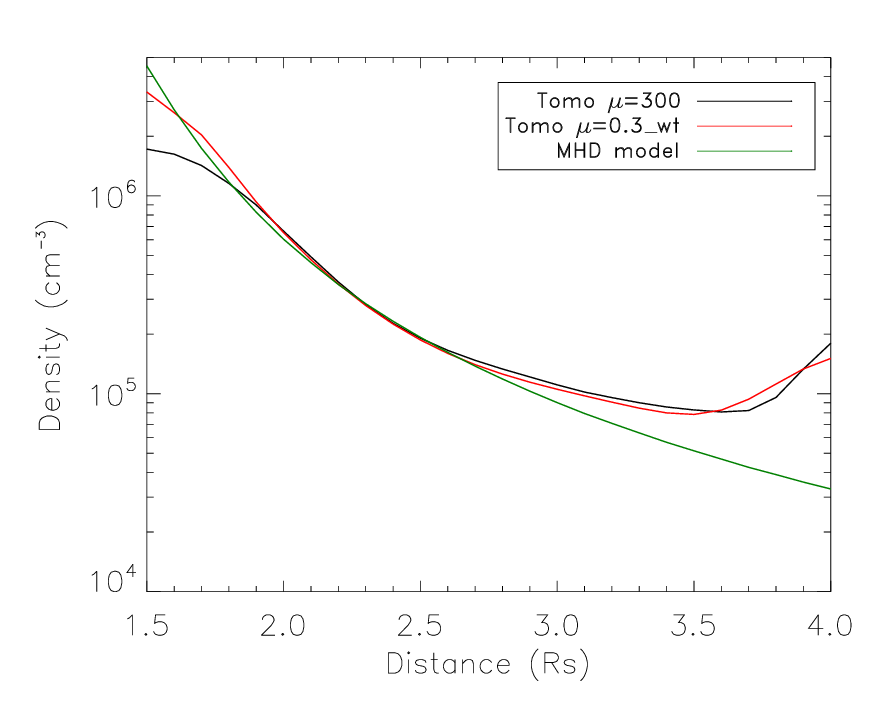}
	    \caption{Radial distributions of the globally-averaged coronal electron density for the tomographic reconstructions without radial weighting (black line), with radial weighting (red line), and from the MHD model (green line).}
            \label{fig:nrcmp}
  \end{SCfigure}

Figure ~\ref{fig:nrcmp} compares the globally averaged radial density profile from the MHD model (green) with those from tomographic reconstructions using non-radial weighting regularization with $\mu=300$ (black) and radial density weighting with $\mu=0.3$ (red). Both tomographic reconstructions display a boundary effect, shown as an increase in density near the outer boundary. This issue arises from the assumption of finite LOS distances and becomes more pronounced close to the boundary. However, the density reconstruction with radial weighting shows a better match to the MHD model near the inner boundary, compared to the non-radial weighting case, which has a peak value (at $r=1.5$\rsun) about twice as high as the latter. 

\subsection{Reconstruction with Zeroth-Order Regularization}  
  \label{sst:zor}
The comparison with the MHD model (see Section~\ref{sst:cmm}) has shown that coronal density structures in the tomographic reconstruction with second-order smoothness appear significantly smoother than the model prediction. Since zeroth-order smoothness is the simplest regularization (see Equation~\ref{eq:reg0}), we examine whether this smoothness can recover more finer density structures. 

\begin{figure}
  \centerline{\includegraphics[width=1.0\textwidth,clip=]{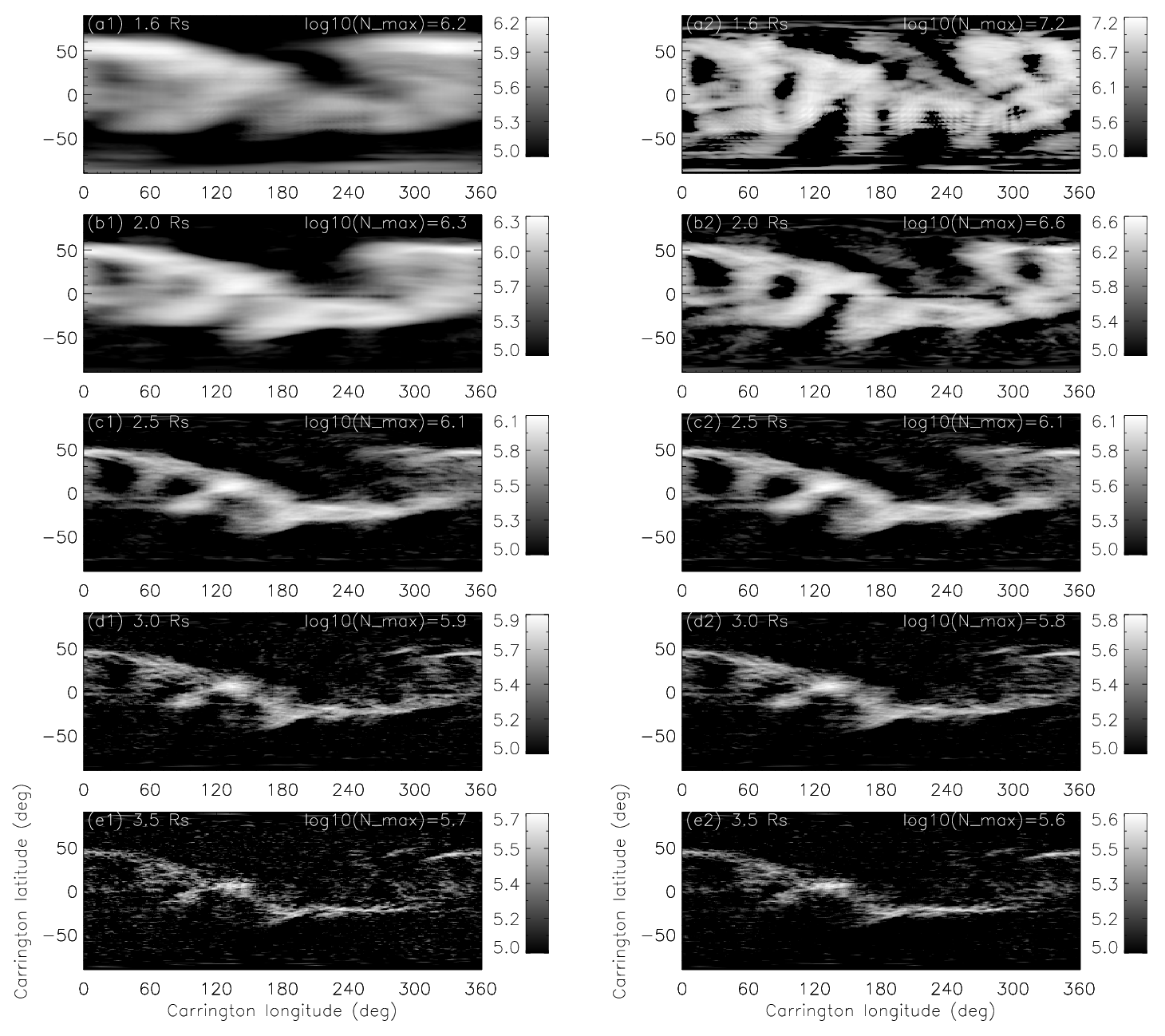}}
  \caption{Comparison between the 3D coronal densities for CR 2098 reconstructed by tomography using zeroth-order regularization without radial weighting ({\it Left panels}) and with radial density weighting ({\it Right panels}). Panels from top to bottom display the density distributions on a logarithmic scale at heights of 1.6, 2.0, 2.5, 3.0, and 3.5 $R_\odot$.}
            \label{fig:nreg0}
  \end{figure}

Figure~\ref{fig:nreg0} shows the 3D coronal density reconstructions for CR 2098 using zeroth-order regularization in two cases: $\mu=5$ without radial weighting (Left panels) and $\mu=0.001$ with radial density weighting (Right panels). Note that the density distributions at each height for the two cases are presented on different logarithmic scales. Panels a1 and a2 show the density distributions at a height of 1.6\,$R_\odot$, rather than 1.5\,$R_\odot$ as in previous cases, because the zeroth-order smoothing failed to recover reasonable density structures due to strong boundary effects.  The boundary effect is more severe in the non-radial weighting case, resulting in a significant density drop near the inner boundary. It is notable that the case with radial weighting recovers finer structures than the case without radial weighting. However, compared to the results using second-order smoothness in both non-radial weighting (see left panels of Figure~\ref{fig:nmap_mbt}) and radial weighting (see left panels of Figure~\ref{fig:ntmd}) cases, using zeroth-order smoothness did not evidently recover finer structures; instead, the reconstructed density structures are noisier, as expected.    

\begin{figure}
  \centerline{\includegraphics[width=1.0\textwidth,clip=]{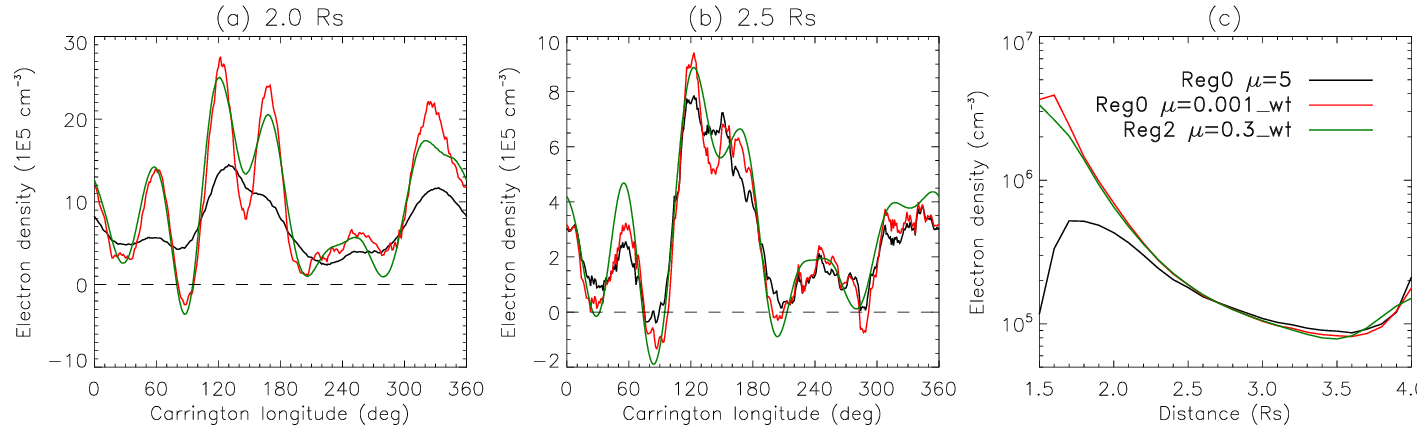}}
\caption{Comparison of density profiles for the CR 2098 reconstructions using zeroth-order regularization without radial weighting (black line) and with radial weighting (red line), and second-order regularization with radial weighting (green line). Density profiles along the equator at (a) 2.0\,$R_\odot$ and (b) 2.5\,$R_\odot$. (c) Radial profiles of the globally-averaged density.}
            \label{fig:nprfreg0}
  \end{figure}

Figure~\ref{fig:nprfreg0} compares the density profiles along the equator at heights of 2.0 and 2.5\rsun\, (panel a and b) and the averaged radial density profile (panel c) for the second-order smoothness case with those for the zeroth-order smoothness cases. At 2.0\rsun, the density profile for the zeroth-order smoothness with radial weighting is comparable to that for the second-order smoothness with radial weighting but has amplitudes about twice as high as the zeroth-order smoothness without radial weighting. However, the density distributions become comparable for the three cases above a height of 2.5\rsun. Figure~\ref{fig:nprfreg0}c clearly indicates that applying radial density weighting resolves the issue of a severe inner boundary effect (significant density drop) in the zeroth-order smoothness reconstruction, resulting in an averaged radial density profile in good agreement with that for the second-order smoothness case.

Other examples comparing reconstructions with second-order smoothing and zeroth-order smoothing can be found in Figure~\ref{fig:nmapevlv} and Figure~\ref{fig:n3view}. These figures consistently show that while the density structures derived from both smoothing methods are similar, the zeroth-order smoothing results display more noise and artifacts. In other words, zeroth-order regularization does not show any advantages in recovering finer density structures compared to higher-order smoothing, and instead introduces more numerical instability.

\subsection{Reconstructions on Spherical Grid}
\label{sst:rsg}
Reconstructions using a spherical grid can avoid artifacts near the inner boundary, which are typically generated when using a Cartesian grid (see Section~\ref{sst:irw}). This is because the spherical grid naturally aligns with the spherical boundaries defined for the solution. 

\begin{figure}
  \centerline{\includegraphics[width=1.0\textwidth,clip=]{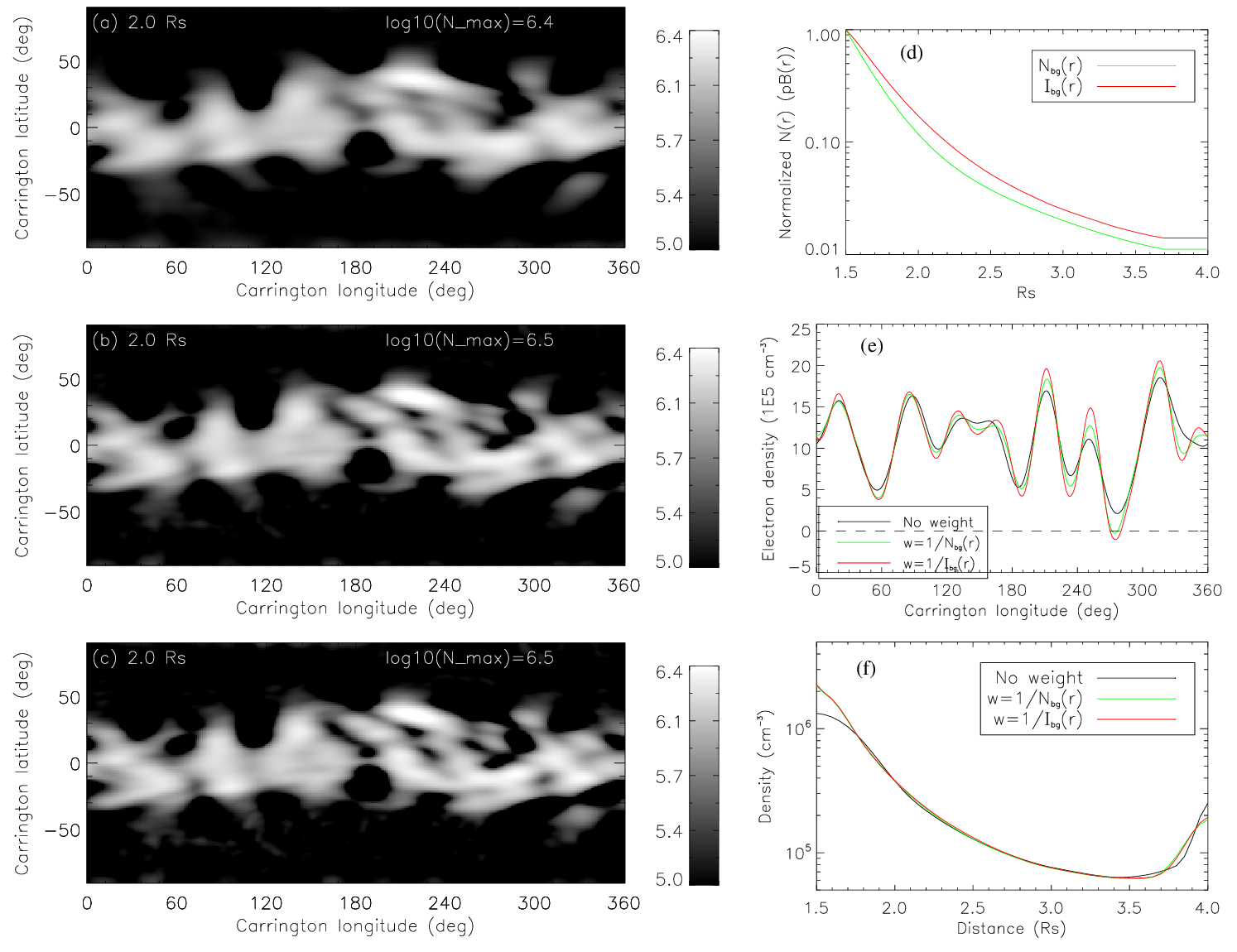}}
	\caption{Reconstructions of 3D coronal electron densities by tomography in spherical grids for CR 2091 in three cases. Density distributions at 2.0\,$R_\odot$ for (a) without radial weighting, (b) with radial density weighting, and (c) with radial pB weighting. (d) Normalized radial profiles of background density (green line) and background pB (red line). (e) Comparison of density profiles along the equator at 2.0\,$R_\odot$ for the three cases. (f) Comparison of radial profiles of globally-averaged density for the three cases. In (e) and (f), the black line represents the case without radial weighting, the green line for the case with radial density weighting, and the red line for the case with radial pB weighting. An animation showing density distributions, as in (a) and (b), at different radial distances is available in the online journal.}
            \label{fig:nsph}
\end{figure}

Figure~\ref{fig:nsph} presents the density reconstructions of CR 2091 using a spherical grid based on pB images accumulated during the first half of its Carrington rotation, from 7 to 20 December 2009. Panel~a shows the density map at 2.0\rsun\, for the case with an optimal regularization parameter of $\mu=40$, determined through CV, with no radial weight applied in regularization. In contrast, panel b and c illustrate two cases where the optimal value of $\mu$ is 0.1 and including radial density weighting ($w=1/N_{bg}(r)$) and pB weighting ($w=1/I_{bg}(r)$), respectively. The normalized density and pB background profiles for these cases are shown in panel d. Panels e and f compare the equatorial density profiles at 2.0\rsun\, and the globally-averaged radial densities for the three cases. 

The comparison reveals that the solutions with radial density and pB weighting are nearly identical. However, the solutions with radial weighting recover finer structures compared to the case without radial weighting. Furthermore, the reconstruction with radial weighting shows improved recovery of densities near the inner boundary, particularly in the range of 1.5$-$1.8\rsun\, (see Figure~\ref{fig:nsph}f), where the peak value at 1.5\rsun\, is approximately 1.7 times higher than in the case without radial weighting. These results are similar to the Cartesian grid case (see Section~\ref{sst:irw}). An animation comparing the density distributions at various heights (from 1.5 to 4.0\rsun) between reconstructions with and without radial weighting is available online.  

\subsection{Effects of Temporal Evolution on Reconstruction}
  \label{sst:uet}
The 3D density reconstructions in this study are based on the assumption that the solution remains static throughout the two-week data collection period from a single satellite. Consequently, the obtained solutions should represent density structures that are either stable or evolving slowly enough to persist over that time. Otherwise, the reconstruction may be interpreted as an average representation of an evolving, dynamic corona, which is likely the case. The truly effective method to overcome the impact of density evolution over time on tomography reconstruction is to use multi-view observational data, which is also an inherent advantage of the tomography technique. To illustrate this impact, we provide two examples below. 

\begin{figure}
  \centerline{\includegraphics[width=1.0\textwidth,clip=]{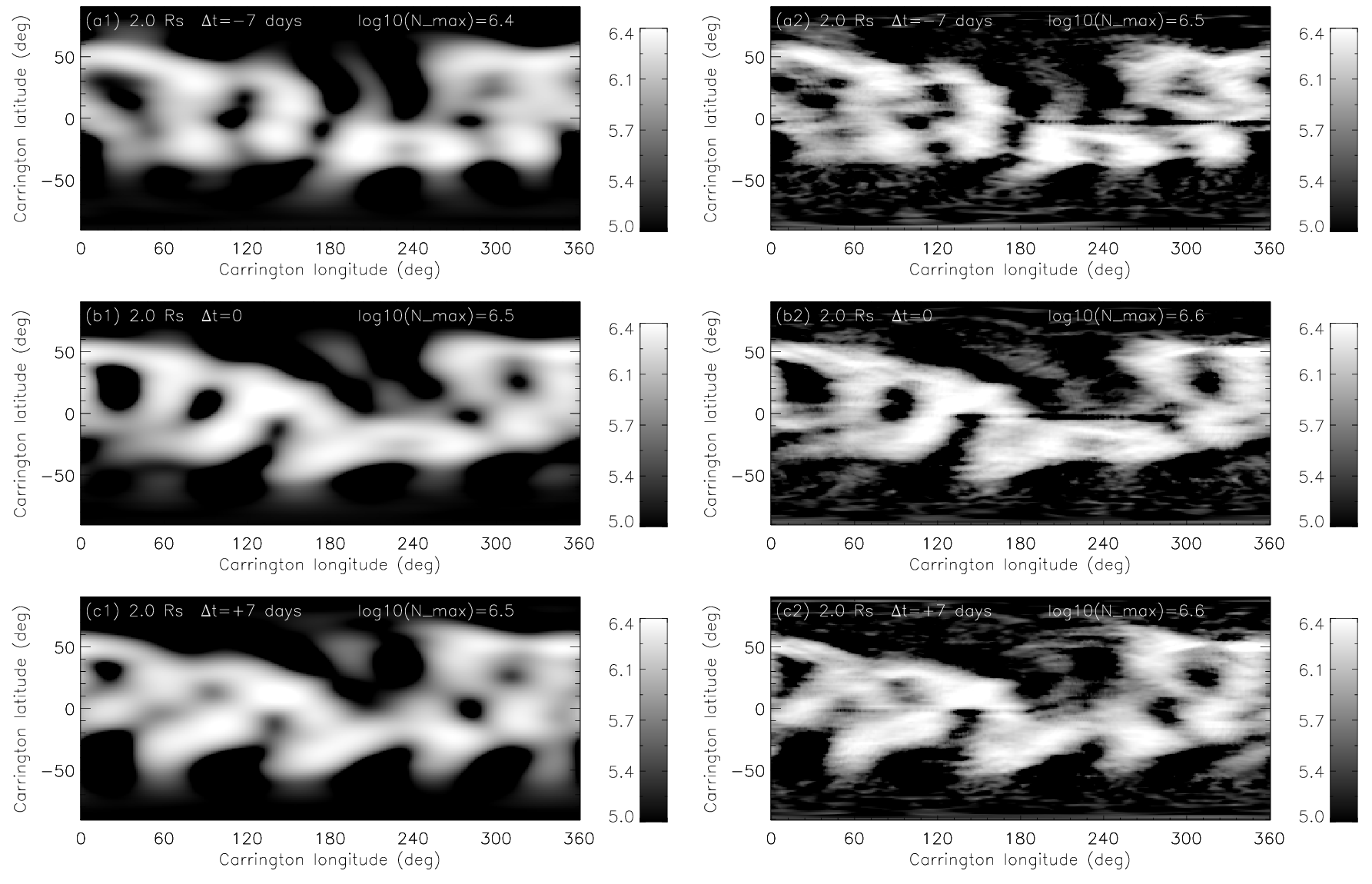}}
\caption{Comparison of 3D coronal electron densities reconstructed for CR 2098 using pB images observed during three different periods: from 23 June to 7 July 2010 ({\it Middle panels}), 7 days earlier from 17 to 30 June 2010 ({\it Top panels}), and 7 days later from 30 June to 13 July 2010 ({\it Bottom panels}). (a1)-(c1): Density distributions at 2.0\,$R_\odot$ for 3D reconstructions derived using second-order regularization with radial density weighting and $\mu=0.3$. (a2)-(c2): Same as (a1)-(c1) but derived using zeroth-order regularization and $\mu=0.001$. All reconstructions are made using a Cartesian grid. An animation showing density distributions, as in (a1)-(c1), at various heights is available in the online journal.}
	    \label{fig:nmapevlv}
\end{figure}

In the first example, we examine how the reconstruction changes with data collection time. Figure~\ref{fig:nmapevlv} compares the reconstruction of CR 2098 (middle panels) with two control experiments: one using data collected a week earlier (upper panels) and the other using data collected a week later (bottom panels). The left panels show the reconstructions derived using second-order smoothing with radial weighting, while the right panels display those using zeroth-order smoothing with radial weighting. The two methods produce results that are in good agreement. An animation of the left panels, displaying the density distributions at various heights (from 1.5 to 4.0\rsun), is available online. The comparison reveals that the primary density structures remain persistent, with only minor changes over time. For example, some differences in the density distributions can be seen around longitude 180$^{\circ}$ between panel a1 and b1, and in the structure at longitudes $60^{\circ}-180^{\circ}$ between panel b1 and c1.  

\begin{figure}
  \centering
  \includegraphics[width=1.0 \textwidth,clip=]{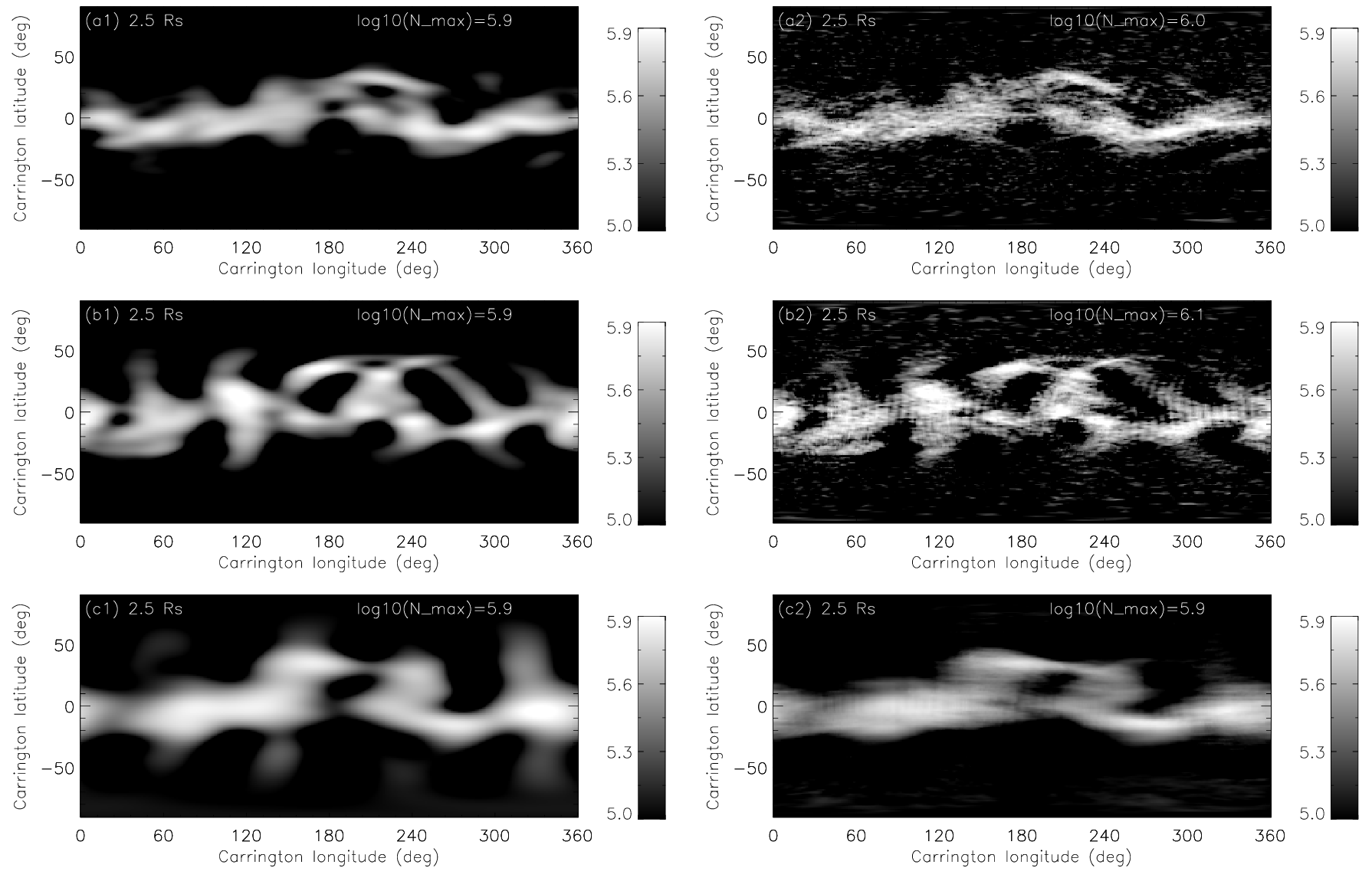}
	\caption{Comparison of 3D coronal electron densities reconstructed for CR 2091 from pB images using single satellite and three satellites. {\it Top panels}: Using observations from STEREO/COR1-B during the first half of CR 2091 (7$-$20 December 2009). {\it Middle panels}: Using STEREO/COR1-B observations from the second half of CR 2091 (21 December 2009 to 3 January 2010). {\it Bottom panels}: Using combined observations from STEREO/COR1-A, COR1-B, and LASCO/C2 during a period around the middle of CR 2091 (18$-$23 December 2009). (a1)-(c1): Density distributions at 2.5\,$R_\odot$ for 3D reconstructions derived using second-order regularization with radial pB weighting where $\mu=0.5$ for single satellite and 1000 for three satellites. (a2)-(c2): Same as (a1)-(c1) but derived using radially-weighted zeroth-order regularization with $\mu=0.002$ for single satellite and 10 for three satellites. All reconstructions are done on a Cartesian grid. An animation showing density distributions, as in (a1)-(c1), at various heights is available in the online journal.}
	      \label{fig:n3view}
\end{figure}

In the second example, we demonstrate a multi-satellite reconstruction of CR 2091, which only requires data collected over approximately five days by leveraging coordinated observations from STEREO/COR1-A, COR1-B, and LASCO/C2. During this period, the three satellites were optimally separated by around 60$^{\circ}$ \citep[see][]{sas19}. The solution was derived within a spherical domain of $2.2\leq{r}\leq 4.0$ \rsun. A detailed description of the method and data processing is provided in \citet{wan23a,wan23b}\footnote{\href{https://agu23.ipostersessions.com/default.aspx?s=81-DE-8F-D4-A3-B0-8D-EC-78-21-50-17-AF-36-15-07&guestview=true}{AGU Fall Meeting 2023, iposter SH03-11}}. Figure~\ref{fig:n3view} compares the density distributions at 2.5\rsun\, from  the three-viewpoint reconstruction of CR 2091 (bottom panels) with two single-viewpoint reconstructions: one using data from the first half CR 2091 (upper panels) and the other using data from the second half CR 2091 (middle panels). Note that the three-viewpoint reconstruction uses data collected around the middle time of CR 2091. The left panels show the reconstructions derived using second-order smoothing with radial pB weighting, while the right panels display those using zeroth-order smoothing with radial pB weighting. Both methods yield consistent results, though the solutions using zeroth-order smoothing appear noisier and exhibit some strip-like artifacts (see panel a2 and b2).

The comparison reveals a distinct difference between the two reconstructions from the first and second halves of CR 2091, indicating that the corona may undergo significant changes over a two-week timescale, even during solar minimum. The clear differences between the three-viewpoint reconstruction and both single-viewpoint reconstructions (e.g., in the structures between longitudes $150^{\circ}-270^{\circ}$) further support this scenario, as the three-viewpoint solution, derived over a shorter timescale, should more accurately reflect the true coronal structure. In other words, if we perform six successive reconstructions for CR 2091 at five-day intervals using multi-viewpoint tomography, we can more reliably study the evolution of the corona.

\section{Discussion and Conclusions}
\label{sct:con}
Tomography has become a vital tool for obtaining more accurate 3D reconstructions of the solar corona without making stringent geometric assumptions, leveraging data from multiple perspectives to build a more complete picture of the electron density structure.

In this study, we improved the regularized tomography technique for reconstructing the 3D coronal density by introducing radial weighting in the regularization term. Tests of the previous method using second-order smoothing based on STEREO/COR1 pB observations show that high-density structures at lower heights (near the inner boundary of the solution domain) are significantly oversmoothed when the regularization parameter is optimally determined by cross-validation. While choosing a smaller regularization parameter can alleviate this issue, it leads to oscillatory artifacts at greater heights. We found that applying radial weighting in the regularization not only overcomes this limitation but also significantly reduces the relative errors of reconstructed densities at higher heights. Radial weighting ensures balanced smoothing of density values across lower and higher heights in the solution. The radial weighting function can be selected as the inverse density background determined from the SSPA density model \citep{wan14} or the inverse intensity background from all pB images used. Comparisons show that reconstructions using radial weighting from both methods are nearly identical. This allows us to choose the latter for simplicity, which is also used as a weighting function in calculating the loss function when solving the minimization problem. 
  
To assess our improved tomography technique, the density reconstruction of CR 2098 obtained from two weeks of STEREO/COR1-B data is compared with the PSI MHD simulation based on the observed magnetic field boundary. Their cross-sectional density distributions at various heights show good agreement, and their radial average density profiles are also highly consistent. The flattening effect of the radial average density profile near the inner boundary, caused by oversmoothing in the case without radial weighting, is effectively corrected. We also compared the density reconstructions of several Carrington Rotations using second-order smoothing and zeroth-order smoothing with radial weighting, and found that their density distributions at different heights are in good agreement. However, without radial weighting, zeroth-order smoothing failed to recover the density near the inner boundary, exhibiting a significant drop with pronounced artifacts. These evaluations confirm that radial weighting is crucial for accurately recovering the 3D coronal density. In addition, comparisons indicate that zeroth-order regularization does not aid in recovering finer density structures compared to second-order smoothing, but instead introduces more numerical artifacts.

We used $n$-fold cross-validation to determine the optimal regularization parameter. The $n$-fold solutions also allow us to estimate the uncertainties in the reconstructed densities due to data sampling. The relative errors are small, typically on the order of a few percent. However, tests that varied the data collection time and compared single-satellite reconstructions with multi-viewpoint reconstructions indicate that coronal evolution over the two-week period has a significant impact. This suggests that the static solution may represent an average of the dynamic corona during the reconstruction period, meaning that only stable, large-scale density structures can be reliably recovered. We demonstrated a multi-viewpoint reconstruction using data from the STEREO/COR1-A, COR1-B, and LASCO/C2 coronagraphs, which required only a five-day data collection period. As a result, the solution is expected to be more reliable and can provide stricter constraints on solar wind models \citep[e.g.][]{sas19, jon22}. 

We developed a tomography inversion code in FORTRAN-90 for both Cartesian and spherical grids. Tests based on STEREO/COR1 data show that the results produced by the two grids are in good agreement. Using this tomography code with radial pB weighting regularization, we have reconstructed 3D densities for CRs 2052 to 2154 (with two reconstructions per CR) using STEREO/COR1-B data, covering the period from 08 January 2007 to 17 September 2014, with a spherical grid of $361\times181\times51$ in longitude, latitude, and radial direction. The dataset of these reconstructions, in the standard FITS format, is publicly available on \href{https://stereo-ssc.nascom.nasa.gov/data/ins_data/secchi/N3D_COR1B}{the SSC webpage}. 

In the future, advanced databases featuring 3D densities reconstructed from multi-viewpoint observations using the improved method will provide a more reliable and accurate depiction of the dynamic solar corona. These databases will help enhance our understanding of the Sun's behavior and contribute to improved models for predicting space weather and solar wind dynamics.

%
 \begin{acks}
	 TW would like to thank Bernd Inhester for instructions and discussions on the tomography algorithm. The authors also appreciate the anonymous referee for constructive comments, which have helped improve the quality of this paper. The STEREO/SECCHI data used here are produced by an international consortium of the Naval Research Laboratory (USA), Lockheed Martin Solar and Astrophysics Lab (USA), NASA Goddard Space Flight Center (USA), Rutherford Appleton Laboratory (UK), University of Birmingham (UK), Max-Planck-Institut f\"ur Sonnensystemforschung (Germany), Centre Spatiale de Li\`ege (Belgium), Institut d’Optique Th\'eorique et Appliqu\'ee (France), Institut d’Astrophysique Spatiale (France).
 \end{acks}

  \begin{authorcontribution}
TW developed the tomography codes on both Cartesian and spherical grids, and conducted all calculations for the results presented in this study, prepared all figures and animations, and wrote the manuscript. CNA and SIJ reviewed the manuscript and offered valuable suggestions and comments. 
  \end{authorcontribution}
  \begin{fundinginformation}
The work of TW and SIJ has been supported by NASA GSFC through Cooperative Agreement 80NSSC21M0180 to Catholic University of America, Partnership for Heliophysics and Space Environment Research (PHaSER). TW also acknowledges support by NASA grants 80NSSC21K1687 and 80NSSC22K0755.
  \end{fundinginformation}
  \begin{dataavailability}
Data sets analyzed and/or generated during the current study are available from the corresponding author on reasonable request.
  \end{dataavailability}

    \begin{codeavailability}
The tomography codes for Cartesian and spherical grids are accessible on \href{https://stereo-ssc.nascom.nasa.gov/data/ins_data/secchi/N3D_COR1B}{the SSC webpage}.
    \end{codeavailability}	    
  \begin{ethics}
  \begin{conflict}
The authors declare that they have no conflicts of interest. 
  \end{conflict}

\noindent	  
{\bf Competing interests}  The authors declare no competing interests.
  \end{ethics}

%
  Using BibTeX
  \bibliographystyle{spr-mp-sola}
  \bibliography{wang.bib}  
%
%
%
%

\end{document}